\documentclass[twocolumn,english,prl, showpacs, superscriptaddress]{revtex4-1}
\usepackage[T1]{fontenc}
\usepackage[utf8]{inputenc}
\usepackage{color}
\usepackage{babel}
\usepackage{amsmath}
\usepackage{amssymb}
\usepackage{graphicx}
\usepackage[unicode=true,pdfusetitle,
 bookmarks=true,bookmarksnumbered=false,bookmarksopen=false,
 breaklinks=false,pdfborder={0 0 1},backref=false,colorlinks=false]
 {hyperref}

\makeatletter

\hypersetup{
    colorlinks=true,
    linkcolor=red,
    citecolor=blue,
    filecolor=blue,      
    urlcolor=blue,
}
\makeatother

\begin{document}
\title{\textcolor{black}{ Thermodynamics of the Ramsey Zone }}
\author{Rogério Jorge de Assis}
\address{Instituto de Física, Universidade Federal de Goiás, 74.001-970, Goiânia
- GO, Brazil}
\address{Departamento de Física, Universidade Federal de São Carlos, 13.565-905,
São Carlos - SP, Brazil}
\author{Ciro Micheletti Diniz}
\address{Departamento de Física, Universidade Federal de São Carlos, 13.565-905,
São Carlos - SP, Brazil}
\author{Celso Jorge Villas-Bôas}
\address{Departamento de Física, Universidade Federal de São Carlos, 13.565-905,
São Carlos - SP, Brazil}
\author{Norton Gomes de Almeida}
\address{Instituto de Física, Universidade Federal de Goiás, 74.001-970, Goiânia
- GO, Brazil}
\begin{abstract}
We carry out a study on thermodynamics properties as entropy and heat $J_{Q}$ and work $J_{W}$ fluxes involved in a Ramsey zone, i.e., a mode field inside a low quality factor cavity that behaves classically and promotes rotations on atomic states. For this, we developed a method to calculate the work associated only with the atom, which would not be possible with the usual approaches that assume a time dependence on the Hamiltonian of the system. Focusing on the atomic dynamic only, here we show that $J_{W}$ predominates when the atomic state evolves maintaining its maximum purity, as computed by von Neumann entropy, in which case the rotation is successfully applied. On the other hand, $J_{Q}$ is the quantity that stands out when the atomic state ceases to be pure due to its entanglement with the cavity field mode state. We describe those limits in terms of the driving strength, the atom-field coupling and the cavity field dissipation rate, and interpret the quantum-to-classical transition in light of the heat and work fluxes. Besides, we show that for a driven-dissipative cavity mode to work out as a Ramsey zone (classical field), a very large amount of photons, of the order of $10^{6}$, need to cross the leaky cavity, which explains the classical behavior of the intra-cavity mode field even though, on average, it has a number of photons of the order of unity {[}Phys.
Rev. Lett. 82, 4737 (1999){]}.
\end{abstract}
\maketitle

\section{Introduction}

Cavity Quantum Electrodynamics (CQED) studies the interaction between
light confined in cavities and atoms where the quantum nature of light
and atoms is significant \citep{Brune1996,Raimond2001,Walther2006}.
In CQED, quantum operations reach the individual control of atomic
levels and their interaction with a single photon for engineering
quantum states, in particular the qubits (two-level systems) that
are now being applied in the construction of quantum computers \citep{Nielsen2011}.

The specific case of a qubit composed by a single two-level atom interacting
with a cavity mode field is described, in the rotating-wave approximation
(valid when the atom-field coupling is much weaker than their natural
oscillation frequency) by the Jaynes-Cummings model \citep{Jaynes1963},
which promotes the so-called Rabi oscillations $\left|g\right\rangle \left|n\right\rangle \leftrightarrow\left|e\right\rangle \left|n+1\right\rangle $
between the atom ground ($\left|g\right\rangle $) and excited ($\left|e\right\rangle $)
states and the cavity state ($\left|m\right\rangle $) through the
interaction term $H_{JC}=\hbar g(\sigma_{+}a+\sigma_{-}a^{\dagger})$,
where $a^{\dagger}$ ($a$) is the creation (annihilation) operator
for the cavity-mode-field and $\sigma_{+}$ ($\sigma_{-}$) is the
raising (lowering) operator for the atom, and $g$ describes the strength
of the Rabi frequency. If the cavity is on resonance with the atomic
transition $\left|g\right\rangle \leftrightarrow\left|e\right\rangle $,
among the fundamental operations between the atom and the field, we
can highlight, for example, the one in which a $\pi/2$ pulse promotes
a coherent exchange of photons between the state of the atomic qubit
and the cavity-mode field qubit, resulting in the evolution $(\alpha\left|g\right\rangle +\beta\left|e\right\rangle )\left|0\right\rangle \leftrightarrow\left|g\right\rangle (\alpha\left|0\right\rangle +\beta\left|1\right\rangle )$.
This type of interaction leaves the field state inside the cavity
in a superposition of vacuum and one-photon states. Likewise, $\pi/4$
pulses starting from $\left|e\right\rangle \left|0\right\rangle $
results in the maximum entangled state $(\left|e\right\rangle \left|0\right\rangle -i \left|g\right\rangle \left|1\right\rangle )/\sqrt{2}$.
On the other hand, atomic superposition states as $\alpha\left|e\right\rangle +\beta\left|g\right\rangle $
can only be obtained using some classical resource, such that the
initial cavity mode $\left|\psi_{i}\right\rangle $ and atomic $\left|a\right\rangle $
states emerge, at the end of the operation, in a product state $\left|a\right\rangle \left|\psi_{i}\right\rangle \rightarrow(\alpha\left|g\right\rangle +\beta\left|e\right\rangle )\left|\psi_{f}\right\rangle $.

\begin{figure}
\begin{centering}
\includegraphics{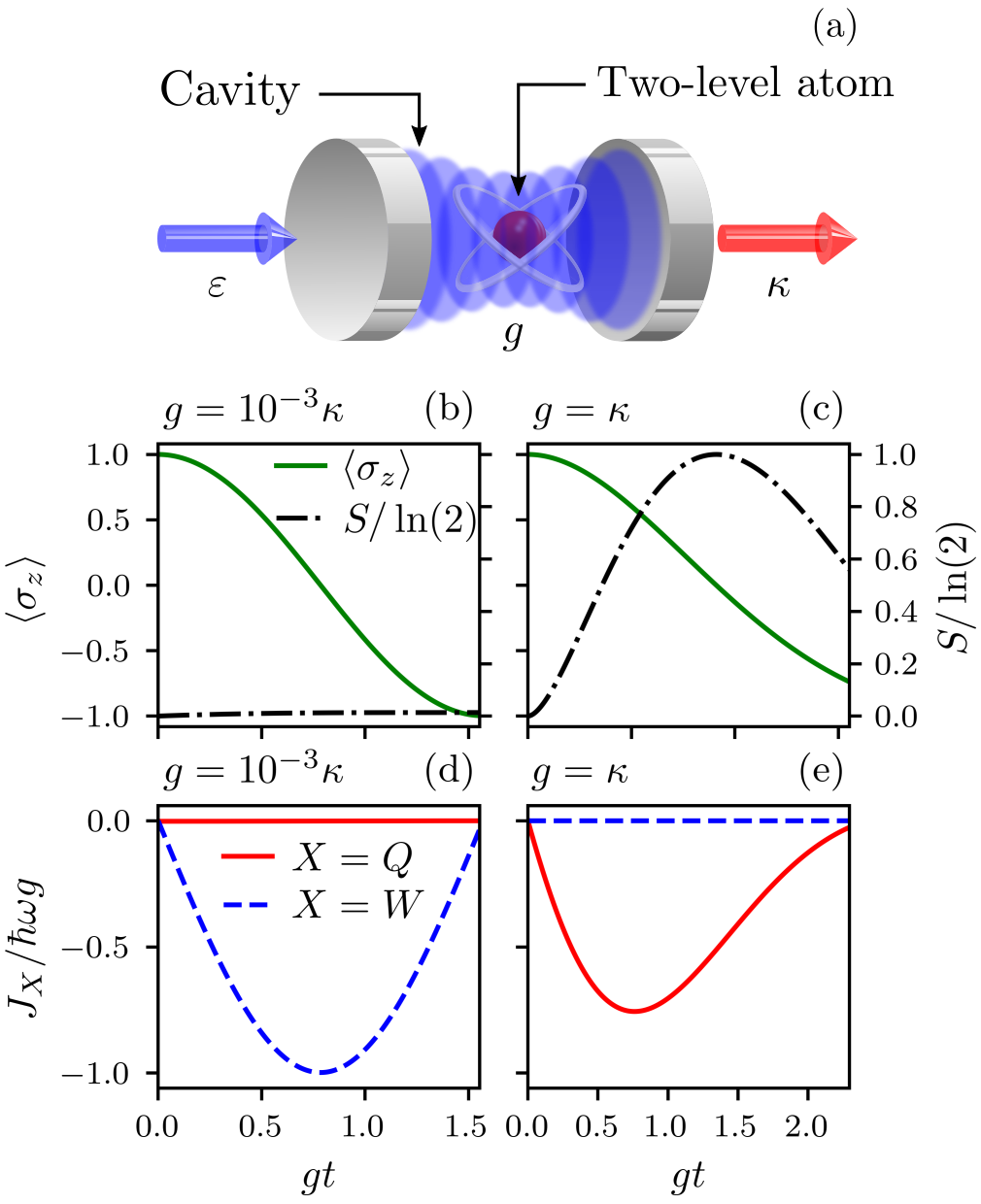}
\par\end{centering}
\centering{}\caption{\label{Fig: 1}Thermodynamics of the Ramsey zone. (a) Experimental
setup, where a two-level atom interacts (coupling $g$) with a single
mode trapped in a low quality factor cavity driven by an external
source, strength $\varepsilon$, and with a decay rate $\kappa$.
Panels (b) and (c) show the atomic population inversion $\langle\sigma_{z}\rangle$
and the normalized von Neumann entropy $S/\ln(2)$ as a function of $gt$. Panels (d) and
(e) show the normalized work flux $J_{W}/\hbar\omega g$ and the normalized
heat flux $J_{Q}/\hbar\omega g$ also as a function of $gt$. Panels
(b) and (d) are for $g=10^{-3}\kappa$ and $\varepsilon=\kappa$,
resulting in null heat flux and non-null work flux during the rotation
process of the atomic state, thus characterizing a unitary evolution.
On the other hand, panels (c) and (e) are for $g=\kappa$ and $\varepsilon=10^{-3}\kappa$,
which results in non-null heat flux and null work flux during the
rotation process of the atomic state, i.e., a purely non-unitary
evolution (when tracing over the mode variables). In all plots we
neglected the atomic decay ($\gamma=0$) and considered $\vert e\rangle\vert\ 0 \rangle$
as the initial atom-field state (in the displaced picture).}
\end{figure}

In the microwave frequency domain, for instance, such superposition of atomic states is generated using the so-called
Ramsey Zone (RZ) \citep{Raimond2001,Walther2006,Hu2018,Kaubruegger2021}, which is schematically illustrated in Fig. \ref{Fig: 1}(a). This technique employs a low quality factor cavity cooled to near absolute zero \citep{Raimond2001,Walther2006,Fink2010}, which is continuously pumped by an external source modeled by $H_{p}=\varepsilon(e^{i\omega_{p}t}a+ e^{-i\omega_{p}t}a^{\dagger})$,
with $\omega_{p}$  and $\varepsilon$ being respectively the frequency and the strength of the driving field, to compensate for the relatively short lifetimes of the photons, such that the cavity
mode field is described by a coherent steady state \citep{Everitt2009,Rossato2011}. Interestingly, even when the low quality factor cavity has one photon on average, thus stressing the quantum character of the cavity-mode field \citep{Kim1999}, the action of the external field, which is the classic resource necessary for the success of this interferometric technique, in addition to the strong cavity field dissipation, produces an effective atom-field interaction that results in a pure rotation on the atomic state only, without the atom-field entanglement that one would expect from purely quantum states.

In this work we study the physics of the Ramsey zone \citep{Kim1999,Rossato2011,Fink2010} from the perspective of the burgeoning field of quantum thermodynamics, which has increasingly attracted the attention of researchers in recent decades \citep{Masanes2017,Perarnau-Llobet2018,Dou2018,Rivas2019,Binder2019,Landi2020,Rivas2020,Strasberg2021}. To this aim, we focus on the atom as the system of interest to quantify
the amount of heat $J_{Q}$ and work $J_{W}$ fluxes involved during
the atom-field-reservoir interaction \citep{Alicki1979, Sai2016}. The purity of the state is quantified by von Neumman entropy, which vanishes for pure states and is maximum for maximally mixed states. Furthermore, von Neumman entropy has the remarkable property of being maximized by Gibbs states, which describe systems in thermodynamic equilibrium. In the present study, where we are dealing with an out-of-equilibrium system, von Neumann entropy provides information about the purity of the atomic state, which is our system of interest.  As we shall
see, while the work is associated directly with the field, a certain amount
of work is also indirectly associated  with the pure atomic rotation, since heat alone, by its
very definition, would not be capable of producing an operation that
results in coherence in the final states. As discussed below, a key ingredient in all of this discussion is a unitary transformation that allows us to work on a displaced picture of the field in the cavity, thus enabling us to easily identify the work and the heat fluxes on the atom. Also, when studying the thermodynamic features of the RZ we will demonstrate that although
on average there is only one photon inside the cavity, for a pure atomic
rotation to take place, it is absolutely necessary that millions of
photons enter and exit the cavity during the time necessary to produce
the desired rotation. This huge amount of photons corresponds to an energy that is absurdly greater
than the work actually needed to produce just the atomic rotation.

In the next section (Sec. \ref{sec:II}) we introduce the model which describes the Ramsey zone and present a method for calculating the work associated with the atom. In Sec. \ref{sec:III} we present
our results concerning the heat and work fluxes \citep{Alicki1979}
in our system and revisit the discussion of how the classical behavior of the field can occur in a cavity that contains on average only one photon. As we will show, for the system to behave classically
it is necessary that a large amount of photons cross the cavity. Finally,
in Sec. \ref{sec:V} we present our conclusions.

\section{\label{sec:II}Model}

The dynamics of a Ramsey zone is described by a master equation composed
of a unitary and a dissipative part \citep{Breuer2002}. The unitary
part is governed by the Jaynes-Cummings and driving field Hamiltonians
($\hbar=1$) \citep{Kim1999,Rossato2011}
\begin{equation}
H=\omega\left(a^{\dagger}a+\frac{1}{2}\right)+\frac{\omega}{2}\sigma_{z}+\big(g\sigma_{+}a+\varepsilon\text{e}^{i\omega t}a+\text{H.c.}\big),
\end{equation}
where the first term describes the cavity mode-field of frequency
$\omega$, the second term describes the two-level atom on-resonance
with the cavity mode field with $\sigma_{z}=\left|e\right\rangle \left\langle e\right|-\left|g\right\rangle \left\langle g\right|$,
the third term describes the atom-cavity mode field interaction with
$g=g^{*}$ being the Rabi frequency, and the fourth term describes
the resonant pumping ($\omega_{p}=\omega$) on the cavity mode field.
This fourth term, as it appears in the equation above, indicates that
the work is associated with the cavity field rather than directly to the atom,
which is our thermodynamic system of interest, and $\text{H.c.}$
stands for the Hermitian conjugate. To obtain a Hamiltonian that reveals
the work that is indirectly associated with the atom, we proceed as follows.
First, we move to a rotating frame according to the interaction picture,
such that $H\rightarrow H_{I}$, with 
\begin{equation}
H_{I}=g\left(\sigma_{+}a+\sigma_{-}a^{\dagger}\right)+\varepsilon \left(a+a^{\dagger}\right).\label{eq: 2}
\end{equation}
Assuming weak interaction between the field and the reservoir modes
and taking into account that experiments are done by cooling the system to near absolute zero, the dynamics of the whole system is governed by
the master equation \citep{Breuer2002} 
\begin{equation}
\dot{\rho}=-i\left[H_{I},\rho\right]+\kappa\mathcal{L}_{a}\left(\rho\right)+\gamma\mathcal{L}_{\sigma_{-}}\left(\rho\right),
\end{equation}
where $\kappa$ ($\gamma$) is the cavity mode field (atom) dissipation
rate, and $\mathcal{L}_{\beta}\left(\rho\right)=2\beta\rho\beta^{\dagger}-\beta^{\dagger}\beta\rho-\rho\beta^{\dagger}\beta$
($\beta=a,\sigma_{-}$) \citep{Kim1999,Rossato2011}. After that,
following Ref. \citep{Rossato2011}, first we write the master equation
in the displaced picture by applying the time-independent unitary
operation $D\left(\alpha\right)=\exp(\alpha a^{\dagger}-\alpha^{*}a)$,
with $\alpha=-i\varepsilon/\kappa$ such that 
\begin{equation}
\dot{\tilde{\rho}}=-i\left[H_{JC}+H_{SC},\tilde{\rho}\right]+\kappa\mathcal{L}_{a}\left(\tilde{\rho}\right)+\gamma\mathcal{L}_{\sigma_{-}}\left(\tilde{\rho}\right),\label{eq: 4}
\end{equation}
with $\widetilde{\rho}=D^{\dagger}\left(\alpha\right)\rho D\left(\alpha\right)$, $H_{JC}=g\left(\sigma_{+}a+\sigma_{-}a^{\dagger}\right)$ and $H_{SC}=\alpha g\sigma_{+}+\alpha^{*}g\sigma_{-}$.
The equation above allows us to identify an effective classical field
driving the atomic state, thus capable to associate work with the atom
\citep{Alicki1979}, and another part that can introduce non-unitarity
to the evolution of the atom, which consists, in addition to the atomic
decay, in interaction with the mode of cavity together with the cavity
dissipation. This can be clearly seen when we deal with effective
dynamics, by tracing over the cavity mode variables. For instance,
considering $g\ll\kappa\sqrt{\left\langle n_{c}\right\rangle +1}$
where $\left\langle n_{c}\right\rangle $ is the intra-cavity average
number of photons (in the displaced picture), we can adiabatically
eliminate the field operators to obtain the effective master equation
to the atom only in the Schrödinger picture \citep{Rossato2011}:
\begin{equation}
\dot{\rho}_{at}=-i\left[H_{at},\rho_{at}\right]+\varGamma_{eff}\mathcal{L}_{\sigma_{-}}\left(\rho_{at}\right),\label{eq: 5}
\end{equation}
where $H_{at}=(\omega/2)\sigma_{z}-(i\varepsilon g/\kappa)(\sigma_{+}\text{e}^{-i\omega t}-\sigma_{-}\text{e}^{i\omega t})$ and $\Gamma_{eff}=g^{2}/\kappa + \gamma$. From this master equation we can directly calculate how much of heat $J_{Q}=\text{tr}(H_{at}\dot{\rho}_{at})$ (dissipative part, which comes from the atom-cavity mode interaction plus the cavity dissipation and the natural atomic decay) and work $J_{W}=tr(\dot{H}_{at}\rho_{at})$ (unitary part) fluxes \citep{Alicki1979} are needed to make the Ramsey zone to work properly. 

Note that for $\gamma =0$ and $g$ small enough, the dynamics is equivalent to that of an atom pumped by a classical field, being approximately unitary and, therefore, only work
is associated with the atom during the evolution. On the other hand, by increasing $g$ the dissipation term can take part in the dynamics, thus allowing the atom to exchange heat with its environment. However, depending on the strength of the driving field $\varepsilon$ one can have either work or heat flux dominating the dynamics (see discussion on the next section.) Furthermore, it is worthwhile
to mention that one of the central points in thermodynamics
is the definition of the system. By defining, therefore, what a system
is and what its surroundings are, we are able to calculate the energy
flows into and out of the system both in the form of heat and work.
In the model we are considering here, the atom is our system of interest,
while the cavity-field, the laser field and the thermal reservoirs
are the surrounding forming the effective thermal bath and external
force for the atom. Having this in mind, note that Eq. \eqref{eq: 4}
must be consistent with the effective master equation (Eq. \eqref{eq: 5}),
thus producing the same result either for heat $J_{Q}$
or work $J_{W}$ fluxes \citep{Alicki1979}. Although useful
for understanding the general aspects of our system, both heat and
work calculated using Eq. \eqref{eq: 5} will be limited in scope
because of the approximation we have made. However, if we now rewrite
Eq. \eqref{eq: 4} as 
\begin{equation}
\dot{\rho}_{at}=-i\left[H_{at},\rho_{at}\right]+\mathcal{L}_{eff}(\rho_{at}),\label{eq: 6}
\end{equation}
where $\dot{\rho}_{at}=\text{tr}_{f}(\dot{\tilde{\rho}})$
and $\mathcal{L}_{eff}(\rho_{at})=\text{tr}_{f}\{-i[H_{JC},\widetilde{\rho}]+\kappa\mathcal{L}(a)\tilde{\rho}+\gamma\mathcal{L}(\sigma_{-})\widetilde{\rho}$,
then we can clearly identify the terms responsible for both heat and
work exchange between the atom and its surrounding. This is an important
remark, since while Eq. \eqref{eq: 5} restrict us to the regime $g\ll\kappa\sqrt{\left\langle n_{c}\right\rangle +1}$,
Eq. \eqref{eq: 6} allows us to numerically investigate heat $J_{Q}$
and work $J_{W}$ fluxes for all values of $g$ and $\varepsilon$.
Here the numerical calculation is done using the quantum optics toolbox
\citep{Johansson2012,Johansson2013} which allows us to easily integrate
the master equation of our system and calculate the desired quantities.

\section{\label{sec:III}Results}

\subsection{Heat and work fluxes in a Ramsey zone}

To investigate how the heat and the work fluxes behave, let's first analyze some extreme cases, where our intuition can help
us to understand our system. To this end, let's assume the initial
state in the displaced picture as $\left|e\right\rangle \left|0\right\rangle $
(i.e., the atom initially prepared in the excited state and
the cavity mode in vacuum, which represents a coherent state that would be reached in the
steady state in the case without atom and in the Schrödinger picture). 

The first case is the one the atom-field coupling $g$ is much smaller than the cavity decay rate, i.e., $g \ll \kappa$, and for appreciable driving strength $\varepsilon$, whose dynamics of the system is governed by Eq. (\ref{eq: 5}). In this case the Jaynes-Cummings dynamics can be ignored, and therefore there will be no entanglement of the atom with the cavity field. This is clearly seen in Fig. \ref{Fig: 1}(b), since the atom performs a complete rotation from the excited to ground state (see the evolution of the $\langle\sigma_{z}\rangle$) keeping the von Neumann entropy $S=-\text{tr}[\rho_{at}\ln(\rho_{at})]$ equals to $0$ during the whole evolution. In this case, as we see in Fig. \ref{Fig: 1}(d), the heat flux $J_{Q}$
is null while the work flux is non-null. Thus, the work flux $J_{W}$
involved in the atomic rotation has been completely directed to successfully
accomplish this task.

In the second case, for intermediate or strong atom-field coupling, i.e., $g \gtrsim \kappa$, and $\left| \varepsilon/\kappa \right| \ll 1$, the Jaynes-Cummings dynamics dominates and the atom-cavity mode state becomes entangled, resulting in a mixed atomic state after tracing over the field variables. Thus, only heat \emph{$J_{Q}$} is exchanged between the atom and its surroundings. This behavior can be seen in Fig. \ref{Fig: 1}(c), where the von Neumann entropy achieves its maximum value ($S=\ln(2)$) during the atomic evolution and the work (heat) flux is null (non-null), as it can be seen in Fig. \ref{Fig: 1}(e).

\begin{figure*}
\begin{centering}
\includegraphics{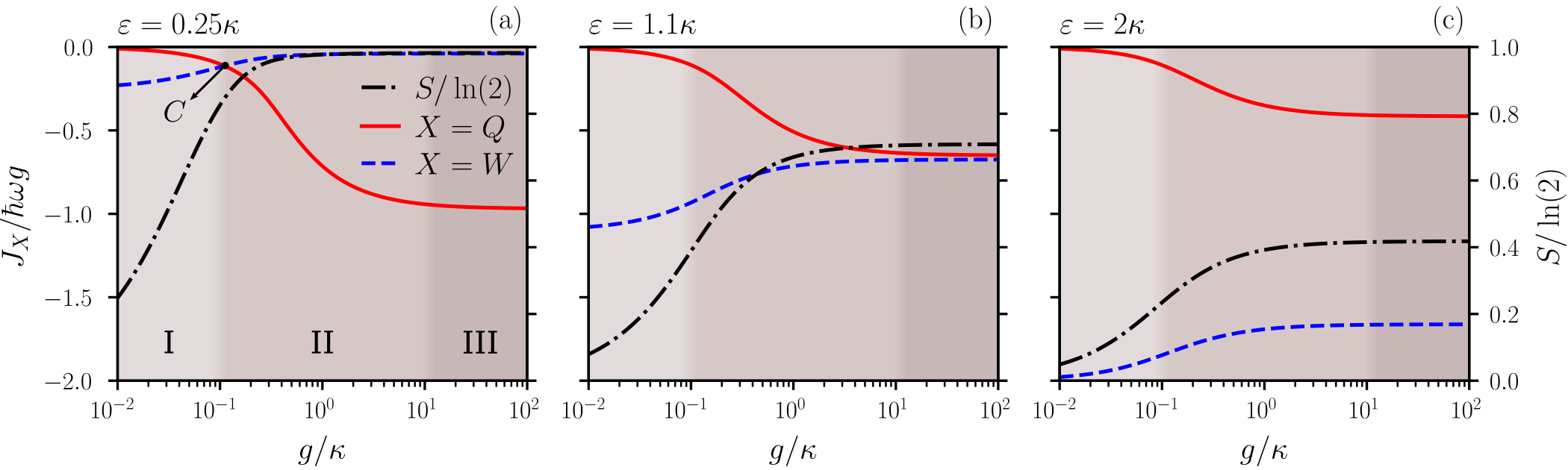}
\par\end{centering}
\centering{}\caption{\label{Fig: 2} 
Normalized von Neumann entropy $S/\ln(2)$ (black dashed-dotted line) and the normalized fluxes of heat $J_{Q}/\hbar\omega g$ (red solid line) and of work on the atom $J_{W}/\hbar\omega g$ (blue dashed line) as a function of the atom-field
coupling $g/\kappa$ for different values of driving field strength:
(a) $\varepsilon/\kappa=0.25$, (b) $\varepsilon/\kappa=1.1$ (critical driving strength), and (c) $\varepsilon/\kappa=2.0$. The heat and work fluxes are computed during the evolution
of the system from the initial state $\left|e\right\rangle \left|0 \right\rangle $
(i.e., the atom initially prepared in the excited state and
the cavity mode in the vacuum state, in the displaced picuture, which represents a coherent state defined by the driving field in the laboratory frame),
until the atom reaches null population inversion, i.e., $\langle\sigma_{z}\rangle=0$.
For all values of $\varepsilon/\kappa$ smaller (bigger) than the
critical point $\varepsilon/\kappa=1.1$ the heat and work fluxes
always cross (do not cross), indicating two different thermodynamic
regimes. For $g \ll \kappa$, the crossing point $C$ is given by Eq. (\ref{Eq: C}). In these plots we also identify tree different regions: I) Region $g \rightarrow 0$, where the atom-field entanglement is negligible and, therefore, the heat flux is also negligible; II) Region where $10^{-1} \lesssim g/\kappa \lesssim 10^1$, in which both normalized heat and work fluxes vary as a function of $g$ (transient region); and  III) Region in that $g \gg \kappa$ and the $J_Q/\hbar g\omega$ and $J_W/\hbar g\omega$ reach finite steady values (as a function of $g$), which depend on the driving strength $\varepsilon$.}
\end{figure*}

Outside of the extreme cases above, i.e., for neither too small
nor too large $\left| g/\kappa \right|$, and intermediate values of  $\left| \varepsilon/\kappa \right|$, the atom can perform work on
the cavity field as well as become entangled with it, thus indicating
that both heat $J_{Q}$ and work $J_{W}$ fluxes are being exchanged
with its surroundings. To analyze the thermodynamics of those cases,
we considered the evolution of the atomic state until it reaches a
population in the excited state $P_e$ equal to the population in the ground
state $P_g$, that is, when $\langle\sigma_{z}\rangle=0$, considering the
atom-field initially in the state $\left|e\right\rangle \left|0\right\rangle $ (in the displaced picture).
Then, at that time, we calculate the heat and work fluxes for the
atom and the von Neumann entropy $S$ for the atomic state. We calculated
those quantities as a function of $g/\kappa$ for different values
of driving strengths $\varepsilon/\kappa$, and the results are shown
in Fig. \ref{Fig: 2}. As expected, for any non null value of
$\varepsilon/\kappa$, the von Neumann entropy starts at minimal values for
$g\rightarrow0$ since, in this case, the influence of the quantum nature of the cavity mode on the atom is negligible (thus resulting
in classical rotations on the atomic state only), and it increases
for $g\rightarrow\infty$. However, its maximum value depends on the
driving strength $\varepsilon/\kappa$: for weak driving the
von Neumann entropy reaches the maximum value $S=\ln(2)$ for stronger atom-field
couplings, as we see in Fig. \ref{Fig: 2}(a), for $\varepsilon/\kappa=0.25$.
This happens because, in this limit, we have basically an atom initially
in the excited state interacting with a cavity mode in the vacuum,
whose final state is a maximally entangled one: $(\left|e\right\rangle \left|0\right\rangle -i\left|g\right\rangle \left|1\right\rangle )/\sqrt{2}$.
On the other hand, for higher values of the atom-field coupling and
stronger driving strengths, the cavity mode reaches a coherent state
(with non null amplitude) which does not lead to a maximally atom-field
entangled state anymore, as we see in Fig. \ref{Fig: 2}(c), $\varepsilon/\kappa=2$.

Focusing on the heat and work fluxes for the atom, we can see that, for a fixed $\varepsilon$, as we increase
the atom-field coupling $g/\kappa$ the normalized heat ($J_{Q}/\hbar g\omega$) and work ($J_{W}/\hbar g\omega$) fluxes always
increases and decreases in modulus, respectively. However, a different thermodynamic
behavior appears depending on the value of the driving strength. For
low (high) values of $\varepsilon/\kappa$ we have (do not have) a
crossing of the heat and work fluxes, as we see in Fig. \ref{Fig: 2}(a)
(Fig. \ref{Fig: 2}(c)) for $\varepsilon/\kappa=0.25$ ($\varepsilon/\kappa=2$).
By solving our system numerically we were able to find out the threshold
between the crossing \textit{versus} no crossing regimes, which happens
for \textbf{$\varepsilon/\kappa\approx1.1$}, as we see in Fig. \ref{Fig: 2}(b).
Thus, for $\varepsilon/\kappa$ below the threshold we always can
find a range of atom-field coupling where the work flux goes to zero
and the heat flux is dominant, thus characterizing a purely quantum
regime (since allows for high degree of atom-field entanglement).
On the other hand, for values of $\varepsilon/\kappa$ above the threshold,
both the work and heat fluxes are non null (with $\left|  J_{Q} \right| < \left| J_{W} \right| $ for
all values of $g/\kappa$), making clear that both quantum and classical
aspects of the cavity field contribute to the atomic dynamics. As discussed above, for $g < \kappa$ the effective dynamics is governed by Eq. (\ref{eq: 5}), which allows us to derive the heat and work fluxes expressions
\begin{equation}
J_{Q}=\text{tr}(H_{at}\dot{\rho}_{at})
= -\frac{2\omega g^2}{\kappa} \langle \sigma_{+}\sigma_{-} \rangle - \frac{2\varepsilon g^3}{\kappa^2}\text{Im}\left[\langle \sigma_{+}\rangle\right], \label{jq}
\end{equation}
\begin{equation}
J_{W}=\text{tr}\left(\dot{H}_{at}\rho_{at}\right) 
= - \frac{2\omega \varepsilon g}{\kappa} \text{Re}\left[\langle \sigma_{+}\rangle\right]. \label{jw}
\end{equation}
Considering the evolution till $\langle \sigma_{+}\sigma_{-} \rangle = P_e =  P_g=1/2$ and keeping only the terms proportional to $\omega$ (since, in the regime we are, $\omega \gg g$), we can find the condition for the crossing point $C$ in Fig. \ref{Fig: 2}(a): $J_Q = J_W$, which reads 
\begin{equation}
    \frac{g}{\varepsilon} =  2\text{Re}\left[\langle \sigma_{+}\rangle\right]. 
    \label{Eq: C}
\end{equation}
For all $g \ll \kappa$ we could numerically verify that $\text{Re}\left[\langle \sigma_{+}\rangle\right] \simeq 0.22$ and then $g \simeq 0.44\varepsilon $ is the point where the heat flux equals the work flux. In Fig. \ref{Fig: 2}(a), for $\varepsilon = 0.25 \kappa$, the crossing point $C$, determined numerically, is given by $g\simeq 0.11\kappa$, in total agreement with Eq. (\ref{Eq: C}). However, for $g \gtrsim \kappa$, Eq. (\ref{eq: 5}) is no longer valid and then the crossing point can not be analytically derived anymore. But, even numerically, it is possible to identify three different regions according to the atom-field coupling which do not depend on the driving field strength (see Fig. \ref{Fig: 2}): I) region in which $g \rightarrow 0$, where the atom-field entanglement is negligible and, therefore, the heat flux is also negligible, and there is a steady behavior for the normalized work flux $J_W/\hbar g\omega$ (as $g \rightarrow 0$). In this region, the value of the normalized work flux depends only on the driving field strength: the higher the $\varepsilon$, the higher the module of normalized work flux $ \left|J_W/\hbar g\omega \right|$; II) region where $10^{-1} \lesssim g/\kappa \lesssim 10^1$, in which both normalized heat and work fluxes vary as a function of $g$ (transient region); and finally III) region in that $g \gg \kappa$ and the $J_Q/\hbar g\omega$ and $J_W/\hbar g\omega$ reach steady values again (as a function of $g$). In this region, both normalized heat and work fluxes can exist, but their values depend again on the strength of the driving field: for $\varepsilon /\kappa \ll 1$, the normalized work flux goes to zero while normalized heat flux becomes non null. On the other hand, for $\varepsilon /\kappa \gg 1$, the normalized heat flux goes to zero while the normalized work flux becomes non null. We can understand the behavior of $J_{W}/\hbar g\omega$ and $J_{Q}/\hbar g\omega$ fluxes in Fig. \ref{Fig: 2}(a)-(c) in terms of entanglement generation. To this end, note from Fig. \ref{Fig: 2}(a) that
when the von Neumann entropy reaches its maximum value, meaning that the atom-mode state approaches its maximum degree of entanglement, $\left| J_{Q}/\hbar g\omega \right|$ increases to its maximum, while $\left| J_{W}/\hbar g\omega \right|$ decreases to its minimum, to the point that $\left| J_{Q}/\hbar g\omega \right|$ becomes greater than $\left| J_{W}/\hbar g\omega \right|$. Interestingly, increasing $\varepsilon$ prevents maximum correlation from forming, as indicated by the von Neumann entropy in Figures \ref{Fig: 2}(b) and (c). As the von Neumann entropy stabilizes at a value lesser than the maximum possible ($S=\ln(2)$), and consequently the correlation stabilizes in a value lower than its maximum possible, the heat generation is also limited, to the point that $\left| J_{Q}/\hbar g\omega \right|$ no longer exceeds $\left| J_{W}/\hbar g\omega \right|$, as indicated by the Figs. \ref{Fig: 2}(b)-(c) in which there is no crossover. Note the dual role of $\varepsilon$: it is responsible for (indirectly) increasing the work performed by the system (atom) and at the same time limits the creation of atom-field entanglement, thereby limiting the amount of heat that flows from the system (atom).

As stated earlier, as our system of interest is only the atom, we
did not conduct a study detailing the work associated with the field or eventually
stored in the atom-field correlations \citep{Modi2015,Huber2015,Alipour2016,Bera2017}.
An investigation to compare the work associated directly with the field with
the work associated indirectly with the atom as well as with the atom-field correlations
could shed some light on the role of the cavity in the extraction process of work from the atom. This will demand more
refined definitions of thermodynamics quantities to describe the exchange
of heat and work involving interacting subsystems evolving under independent
reservoirs. This is an interesting point and will be investigated
in the future.

\subsection{\label{subsec: B}On the classical behavior of the cavity field }

\begin{figure}
\begin{centering}
\includegraphics{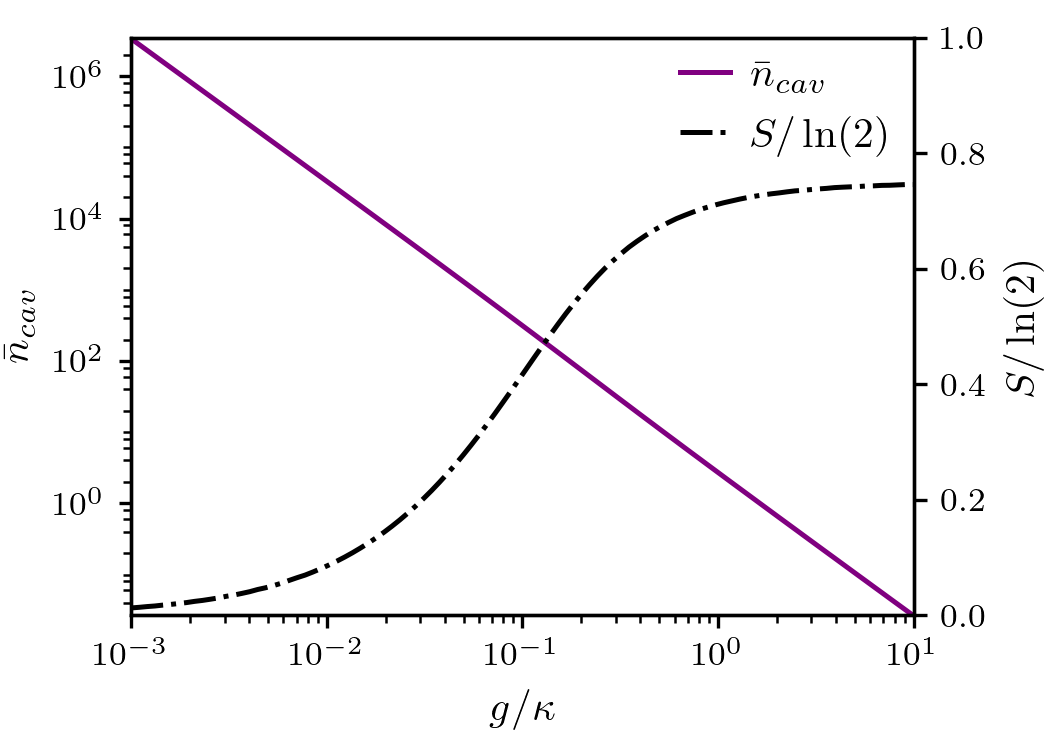}
\par\end{centering}
\caption{\label{Fig: 3}Normalized von Neumann entropy $S/\ln(2)$ (black dashed-dotted line) and average photon number
$\bar{n}_{flux}$ (purple solid line) crossing the lossy cavity \emph{versus}
$g/\kappa$ at time when the atom achieve the state for $P_{e}=\left\langle e\right|\rho_{at}\left|e\right\rangle \simeq P_{g}=\left\langle g \right|\rho_{at}\left|g\right\rangle \approx0.5$,
i.e., the population of the ground and excited states are the
same for fixed $\bar{n}_{cav}=\left| \varepsilon/\kappa\right| ^2=1$, which is the
average number actually found into the cavity.}
\end{figure}

In this Subsection we discuss an important issue raised in Ref. \citep{Kim1999},
which is the fact that one can have, on average, only one photon inside
the cavity and, even so, the cavity mode field can be treated from
a classical perspective, since the atom-field entanglement that would
be expected between two quantum systems does not occur. As we see in Fig. \ref{Fig: 2}, this happens when the heat flux becomes  negligible, i.e., when $g \ll \varepsilon$.  To address this point more carefully, first note
that during the time of unitary operation, the pumping term $H_{p}=\varepsilon(a^{\dagger}+a)$
is responsible to take the cavity state from $\left|0\right\rangle $
to $\left|\alpha(t)\right\rangle =\exp\left[-i\varepsilon t(a^{\dagger}+a)\right]\left|0\right\rangle $,
with $\alpha=-i\varepsilon t$ in Eq. \eqref{eq: 2}. Therefore, the
amount of photons that cross the cavity while the unitary operation
is taking place is given by $\bar{n}_{flux}=\left|\varepsilon t\right|^{2}$.
On the other hand, the number of photons that actually remain inside
the cavity is given by the difference between what is pumped into
the cavity and what leaks through the walls of the low quality cavity,
i.e., $\bar{n}_{cav}=\left|\varepsilon/\kappa\right|^{2}$,
resulting in $\bar{n}_{flux}=\kappa^{2}t^{2}\bar{n}_{cav}$.
Therefore if we require $\bar{n}_{cav}\sim1$, then $\bar{n}_{flux}\sim k^{2}t^{2}$.
To the Ramsey zone work properly, the $\kappa\gg g$ limit must be
met. For cavities with a low quality factor \citep{Raimond2001,Walther2006},
we can estimate not only how many photons actually go trough the cavity
during the time the atom crosses the cavity, but how much energy $E=\hbar\omega n_{flux}$
must be invested for the unit operation $U$ (i.e, the Ramsey
zone) to be performed successfully. Now, if we note that a typical
atom rotation occurs for $\varepsilon gt/\kappa\sim\pi$ then
$\bar{n}_{flux}\sim\kappa^{2}/g^{2}$, and we can therefore estimate,
assuming $\kappa/g\sim10^{3}$, a total number of photons crossing
the low Q cavity as $\bar{n}_{flux}\sim10^{6}$ photons. This huge
amount of photons helps to understand the emergence of the classical
behavior even when there is only one photon on average inside the
cavity. As noted in Ref. \citep{Kim1999}, there is nothing special
about choosing $\bar{n}_{cav}=\left|\varepsilon/\kappa\right|^{2}\sim1$:
the unitary operation can be accomplished even for $\bar{n}_{cav}=\left|\varepsilon/\kappa\right|^{2} \gg 1$,
provided the requirement $\kappa \gg g$ is met. However, this would
require to increase the velocity of the atom crossing the cavity.
This large expenditure of energy $\bar{E}_{flux}=10^{6}\hbar\omega$
is in contrast to the energy $\bar{E}_{cav}$ that is actually needed
to produce work on the cavity field, as shown in Fig. \ref{Fig: 2}.
To better appreciate how the total average number of photons inside
the cavity changes with the parameters involved, in Fig. \ref{Fig: 3}
we show, in a log scale on the left, the number of photons crossing
the cavity as a function of the $g/\kappa$ (purple solid line) for the evolution from  $\langle \sigma_z \rangle \simeq 1$ till  $\langle \sigma_z \rangle \simeq 0$
 ($\pi/2$-rotation), which would represent, in a perfectly unitary evolution, to a superposition state $(\left|e\right\rangle-i\left|g\right\rangle)/\sqrt{2}$. The von Neumann entropy (black dashed-dotted line) is also shown on the left scale.
Note that to produce with high purity a rotation of $\pi/2$, i.e., with von Neumann entropy close to zero, it is necessary a high number
of photons crossing the cavity, on the order of $10^{6}$.

\section{\label{sec:V}Conclusion}

In this work, using a method we developed, we studied the thermodynamic quantities entropy $S$, heat $J_{Q}$
and work $J_{W}$ fluxes exchanged by an atom and its environment
in the functioning of a Ramsey zone, which is a device employed to rotate atomic
qubits. It is constituted by a field, which interacts with the atom, in a lossy cavity whose energy is kept constant due to a pumping field resonant with the atom and the cavity mode. We show that for parameters for which the Ramsey zone works
properly, i.e., without entanglement generation and therefore causing
the atomic state to evolve with a high degree of purity (minimal entropy) $J_{W}$
is the amount that stands out. On the other hand, for parameters for
which the Ramsey zone fails to function, the atom state becomes highly
entangled with the cavity mode field state, and therefore there is
a drastic decrease in the purity of the atomic state, as shown by the von Neumann entropy, in which case
the amount that stands out is $J_{Q}$. Yet, for certain parameters where
the degree of entanglement between the atom state and the cavity mode
field state is finite but not maximum, and therefore the von Neumann entropy is neither minimum 
nor maximum, both heat and work are present. In addition, we demonstrate the existence of a specific value for $\varepsilon$ beyond which entanglement generation no longer occurs, and consequently no more heat can be extracted from the system (atom). Furthermore, our study
reveals that the average amount of photons coming from the classical
pump and crossing the lossy cavity during the Ramsey zone operation
is of the order of millions of photons, which explains the classical
behavior of the cavity mode, even if on average the lossy cavity contains
only one photon. The results presented here provides a way to understand the dynamics of the Ramsey zone through quantum thermodynamics concepts, being of interest to the quantum optical and thermodynamics community. In particular, the method we developed allowed us to calculate both the work and the heat associated with the atom, which would not be possible using usual approaches applicable only to time-dependent Hamiltonians. This means that our method has potential application in quantum thermal machines that use the concepts of heat and work as a figure of merit for the calculation of efficiency and performance.

\section*{Acknowledgments }

We acknowledge financial support from the Brazilian agencies: Coordenação de Aperfeiçoamento de Pessoal de Nível Superior (CAPES), Financial code 001, National Council for Scientific and Technological Development (CNPq), Grants No. 311612/2021-0 and 301500/2018-5, São
Paulo Research Foundation (FAPESP) Grants No. 2019/11999-5 and No. 2021/04672-0, and Goiás State Research Support Foundation (FAPEG).
This work was performed as part of the Brazilian National Institute
of Science and Technology for Quantum Information (INCT-IQ/CNPq) Grant
No. 465469/2014-0. C.M.D thanks CNPq for financial support.

\bibliographystyle{apsrev4-1}
\bibliography{References}

\end{document}